\documentclass[twocolumn,showpacs,preprintnumbers,nofootinbib,amsmath,amssymb]{revtex4}

\usepackage{euscript,amssymb,amsmath}

\usepackage{graphicx}
\usepackage{epsfig}

\usepackage{hyperref}
\usepackage{color}
\usepackage[normalem]{ulem}

\newcommand{\be}[1]{\begin{equation}\label{#1}}
\newcommand{\ee}{\end{equation}}
\newcommand{\ba}[1]{\begin{eqnarray}\label{#1}}
\newcommand{\ea}{\end{eqnarray}}
\newcommand{\rf}[1]{(\ref{#1})}
\newcommand{\nn}{\nonumber}

\newcommand{\td}{\tilde}

\newcommand{\de}{\partial}

\newcommand{\om}{\omega}

\newcommand{\ve}{\varepsilon}

\newcommand{\La}{\Lambda_5}

\begin{document}

\title{Brane world models with bulk perfect fluid and broken 4D Poincar$\mathbf{\acute{e}}$ invariance}

\author{\"{O}zg\"{u}r Akarsu$^a$}\email{akarsuo@itu.edu.tr} \author{Alexey Chopovsky$^b$}\email{a.chopovsky@yandex.ru}
\author{Maxim Eingorn$^c$}\email{maxim.eingorn@gmail.com} \author{Seyed Hossein Fakhr$^b$}\email{seyed.hossein.fakhr@gmail.com}
\author{Alexander Zhuk$^{a,b}$}\email{ai.zhuk2@gmail.com}

\affiliation{$^a$Department of Physics, \.{I}stanbul Technical University, 34469 Maslak, \.{I}stanbul, Turkey} \affiliation{$^b$Astronomical Observatory, Odessa
National University, 2 Dvoryanskaya st., Odessa 65082, Ukraine} \affiliation{$^c$North Carolina Central University, CREST and NASA Research Centers, 1801
Fayetteville st., Durham, NC 27707, U.S.A.}

\begin{abstract}
We consider 5D brane world models with broken global 4D Poincar\a'{e} invariance (4D part of the spacetime metric is not conformal to the Minkowski spacetime).
The bulk is filled with the negative cosmological constant and may contain a perfect fluid. In the case of empty bulk (the perfect fluid is absent), it is shown that one brane solution always has either a physical or a coordinate singularity in the bulk. We cut off these singularities in the case of compact two brane model and obtain regular exact solutions for both 4D Poincar\a'{e} broken and restored invariance. When the perfect fluid is present in the bulk, we get the master equation for the metric coefficients in the case of arbitrary bulk perfect fluid equation of state (EoS) parameters. In two
particular cases of EoS, we obtain the analytic solutions for thin and thick branes. First one generalizes the well known Randall-Sundrum model with one brane to
the case of the bulk anisotropic perfect fluid. In the second solution, the 4D Poincar\a'{e} invariance is restored. Here, the spacetime goes asymptotically to
the anti-de Sitter one far from the thick brane.
\end{abstract}


\maketitle

%%%%%%%%%%%%%%%%%%%%%%%%%%%%%%%%%%%%%%%%%%%%%%%%%%%%%%%%%%%%%%%%%

\section{\label{sec:1}Introduction}

\setcounter{equation}{0}

The idea on multidimensionality of our spacetime has more than hundred years of history starting from the pioneering paper by Nordstr\"{o}m
\cite{Nordstrom:1988fi} published in 1914 just before the Einstein's General Relativity came to light. A brief historical retrospective of the main ideas and
papers as well as introduction and review papers and books can be found in the recent article \cite{Liu:2017gcn}. The modern turn of the history is connected with
the so called brane world models. Here, it is supposed that our visible world is localized on 4D hypersurface (brane) embedded in multidimensional spacetime
(bulk). The progress of these models was largely due to the papers by Lisa Randall and Raman Sundrum \cite{Randall:1999ee,Randall:1999vf} who proposed a
non-factorizable warped geometry for solving the gauge hierarchy problem. This idea caused an ongoing flood of articles devoted to the study of properties and
various modifications of this geometry (see, e.g., reviews pointed out in \cite{Liu:2017gcn}).

In original papers by Randall and Sundrum as well as in most subsequent articles, it was proposed that metric is Poincar\a'{e} invariant. This means that 4D
part/section of metric is conformal to the Minkowski spacetime (see, e.g., \cite{ArkaniHamed:2000eg,Kachru:2000hf}). The common conformal prefactor in front of
the 4D Minkowski metric depends on the extra dimension (and time in the case of cosmological implementation). However, what new properties will brane world models
exhibit if we violate such global conformal connection and suppose that the temporal and 3D spatial parts have different prefactors? In this case the global 4D
Poincar\a'{e} invariance is broken\footnote{Obviously, for each 4D section we can restore the local Poincar\a'{e} invariance with the help of redefinition of the
time coordinate. However, for different sections this redefinition will be different.}.  This is the main subject of our studies in the present paper.

We consider the static 5D metric with the broken global 4D Poincar\a'{e} invariance. In general, we do not require the reflection $\mathbb{Z}_2$ symmetry
for the metric coefficients. Such asymmetric brane solutions were investigated, e.g., in the papers \cite{Ida:1999ui,Battye:2001yn,Gergely:2003pn,Maia:2004fq} where the metric ansatz was taken in
the form of the five-dimensional analogue of the Schwarzschild-anti-de Sitter spacetime \cite{Ida:1999ui,Battye:2001yn,Gergely:2003pn}. We consider a different metric ansatz. We also fill
bulk with the negative cosmological constant and perfect fluid with anisotropic equations of state (EoS).
%It turns out

The results of our investigations are twofold. First, we demonstrate that the behavior of models with broken and restored invariance is significantly different
from each other. Second, this setting of the problem enables us to obtain new classes of solutions. For example, in the case of the empty bulk (the perfect fluid is absent) the solution  always has the singularity (naked or coordinate) in contrast to the usual Poincar\a'{e} invariant models (e.g., \cite{Randall:1999ee,Randall:1999vf}). Such type of naked singularities is known for the models with the bulk scalar field and restored Poincar\a'{e} invariance
\cite{ArkaniHamed:2000eg,Kachru:2000hf,Cohen:1999ia,Forste:2000ps,Forste:2000ft}. In these papers, the singularities are treated as Big Bang or Big Crunch and
they are taken to effectively cut off space. However, we prefer to construct completely regular solutions. Therefore, we introduce a second brane which cuts off
the singular points where the metric coefficients either are infinite or equal to zero. We find the range of parameters which ensures such regular solutions
defined on the compact space.

In the presence of the perfect fluid in bulk, the system of equations is reduced to one master equation for the metric coefficients. In the case of arbitrary EoS,
this equation does not allow to get analytic expressions for the metric coefficients. However, this equation is useful for numerical studies of the considered
brane world models. We present two physically interesting particular analytic solutions for the metric coefficients. The first one generalizes the Randall-Sundrum
solution with one brane (RSII) \cite{Randall:1999vf} to the case of broken Poincar\a'{e} invariance and bulk anisotropic perfect fluid. The second analytic
solution describes the thick brane with the anisotropic bulk perfect fluid and restored Poincar\a'{e} invariance. This solution is of interest since far from the
thick brane it goes asymptotically to the anti-de Sitter one. As far as we are aware, the thick brane solutions were constructed mainly for the brane world models
with the bulk scalar field \cite{Liu:2017gcn,Dzhunushaliev:2009va}. The only known for us thick brane solution with a perfect fluid is presented in the paper
\cite{Ivashchuk:2001zd}. However, the exact analytic expressions for the metric coefficients are not given in this article.

The paper is structured as follows. In Sec. \ref{sec:model}, the general setting of the model is given. In Sec. \ref{sec:eb}, we consider the empty bulk model.
Here, in subsections \ref{subsec:A} and \ref{subsec:B}, we consider one- and two-brane models, respectively. Sec. \ref{sec:bulk} devoted to the models with
perfect fluids in bulk. The case of arbitrary perfect fluid equations of state (except for a couple of special cases) is considered in subsection \ref{subsec:4A}.
Here, we obtain a master equation for the metric coefficients. In subsections \ref{subsec:4B} and \ref{subsec:4C}, we obtain analytic expressions for the metric
coefficients in some particular cases of EoS parameters. For example, the case of subsection \ref{subsec:4C} describes the thick brane solution. The main results
are summarized in concluding Sec. \ref{sec:conc}.

%%%%%%%%%%%%%%%%%%%%%%%%%%%%%%%%%%%%%%%%%%%%%%%%%%%%%%%%%%%%%%%%%%%%%%%%%%%%%%%%%%%
\section{The model}
\label{sec:model}

\setcounter{equation}{0}

We  consider the static 5D metric in the form
%%%%%%%
\be{2.1}
{\rm d}s^2=A(\xi){\rm d}t^2+B(\xi)({\rm d}x^2+{\rm d}y^2+{\rm d}z^2)+E(\xi){\rm d}\xi^2\, .
\ee
%%%%%%%
Obviously, without loss of generality, we can put $E(\xi)\equiv -1$. In the case $A(\xi)=-B(\xi)$, 4D sections (e.g., branes) $\xi=\mathrm{const}$ are
Poincar\a'{e} invariant. However, in our work we do not require such invariance letting functions $A(\xi)$ and $B(\xi)$ to be arbitrary. The bulk is filled with
the negative cosmological constant\footnote{In the present paper we consider the case $\Lambda_5<0$ since the negative bulk cosmological constant is a natural
feature of the string theory or M-theory, e.g., in this case we can introduce the AdS/CFT correspondence \cite{Maldacena:1997re}. However, the case $\Lambda_5>0$
is not forbidden in our model and can be studied in the similar way.} $\Lambda_5<0$ and a perfect fluid with mixed energy-momentum tensor components
%%%%%%%%
\be{2.2}
T_{0}^{0}=\varepsilon,\quad T_{1}^{1}=T_{2}^{2}=T_{3}^{3}=-p_{0},\quad T_{4}^{4}=-p_{1}\, .
\ee
%%%%%%%%
For such a model, 5D Einstein field equations are reduced to the system of three equations:
%%%%%%
\ba{2.3}
&{}&-\frac{3B''}{2B}-\kappa\Lambda_{5}=\kappa\varepsilon\, ,\\
\label{2.4}
&{}&\frac{B''}{B}+\frac{A'B'}{2AB}-\frac{(B')^{2}}{4B^2}+\frac{A''}{2A}-\frac{(A')^{2}}{4A^{2}}+\kappa\Lambda_{5}\nn \\
&{}&=\kappa p_0=\kappa\, \omega_0\varepsilon\, ,\\
\label{2.5}
&{}&\frac{3(B')^2}{4B^2}+\frac{3A'B'}{4AB}+\kappa\Lambda_{5}=\kappa p_1=\kappa\, \omega_1 \varepsilon\, ,
\ea
%%%%%%
where $\kappa\equiv 2\pi^2G_5/c^4$ with $G_5$ being the gravitational constant in the 5-dimensional spacetime, $'$ stands for the derivative with respect to $\xi$
and we assumed that the EoS are of the form $p_0=\omega_0\varepsilon$ and $p_1=\omega_1\varepsilon$.
%%%%%%%%%%%%%%%%%%%%%%%%%%%%%%%%%%%%%%%%%%%%%%%%%%%%%%%%%%%%%%%%%%%%%%%%%%%%%%%%%%%%

%%%%%%%%%%%%%%%%%%%%%%%%%%%%%%%%%%%%%%%%%%%%%%%%%%%%%%%%%%%%%%%%%%%%%%%%%%%%%%%%%%%
\section{Empty bulk}
\label{sec:eb}

\subsection{One-brane model}
\label{subsec:A}

Let us study first the case of empty bulk: $\varepsilon\equiv 0$. In this case, Eqs. \rf{2.3}, \rf{2.4} and \rf{2.5} read
%%%%%%%
\ba{3.1}
&{}&\cfrac{B''}{B}=-\cfrac{2}{3}\,\La\, ,\\
\label{3.2}
&-&\cfrac{B''}{B}-\cfrac{1}{2}\,\cfrac{A''}{A}+\cfrac{1}{4}\cfrac{B'^2}{B^2}-\cfrac{1}{2}\cfrac{A'B'}{AB}+\cfrac{1}{4}\cfrac{A'^2}{A^2} =\La \, ,\\
\label{3.3} &{}&\cfrac{B'^2}{B^2}+\cfrac{A'B'}{AB}=-\cfrac{4}{3}\,\La\, ,\ea
%%%%%%%
where we put for a moment $\kappa\equiv 1$. It can be easily seen that if Eqs. \rf{3.1} and \rf{3.3} are satisfied, then Eq. \rf{3.2} is satisfied automatically.
Therefore, it is sufficient to solve only \rf{3.1} and \rf{3.3} together. The general solution of these equations is
%%%%%%%%
\ba{3.4}
B(\xi)&=&B_1e^{m\xi}+B_2e^{-m\xi},\,\, m^2\equiv-\cfrac{2}{3}\,\La>0\, ,\\
\label{3.5}
A(\xi)&=&A_1\,\cfrac{[B'(\xi)]^2}{B(\xi)}\, ,
\ea
%%%%%%%
where $B_1, B_2$ and $A_1$ are arbitrary constants of integration.

We can now put a brane, e.g., at the point $\xi=0$ requiring the $S^1/\mathbb{Z}_2$ symmetry with respect to this point. The well known Randall-Sundrum solution
(RSII) \cite{Randall:1999vf} corresponds to the additional condition $A(\xi )=-B(\xi)$ which restores the 4D Poincar\a'{e} invariance. This condition results in
either $B_1=0$ or $B_2=0$ (in the case of the original solution \cite{Randall:1999vf} we should substitute $B_1=0$ for a positive value of $m$) and $A_1=-1/m$. In
this case we obtain a regular solution at any point $\xi \in [0, \pm\infty)$. However, in general case $A(\xi)\neq -B(\xi)$ there are singular points where
$B(\xi)=0$. For example, the Kretschmann invariant for the metric \rf{2.1} is
%%%%%%
\ba{3.6}
\hspace*{-1cm}&{}&K=R^{MNKL}R_{MNKL}=\frac{A''^2}{A^2}+\frac{3 A'^2
   B'^2}{4 A^2
   B^2}\nn \\
\hspace*{-1cm}&+&\frac{A'^4}{4 A^4}
   -\frac{A'^2
   A''}{A^3}
   +\frac{3 B''^2}{B^2}+\frac{3
   B'^4}{2 B^4}-\frac{3 B'^2
   B''}{B^3},
\ea
%%%%%%
and for the solutions \rf{3.4} and \rf{3.5} it reduces to
%%%%%%
\be{3.7}
K=   \frac{m^4}{2}
   \left[\frac{(12 B_1
  B_2)^2}{\left[B(\xi)\right]^4}+5\right]\, .
\ee
%%%%%%
Therefore, it diverges when $B(\xi)=0$. Hence, physical singularities are localized at the points where $B(\xi)=0$, while at the points where $A(\xi)=0$ the Kretschmann invariant shows regular behaviour. In the next section (see the text after Eq. \rf{3.14}), we demonstrate that in the case of the 4D Poincar\a'{e} invariance violation the singular points with the physical singularity (where $B(\xi)=0$ and we have the curvature singularity) or the coordinate singularity (where $A(\xi)=0$) always exist. Therefore, the empty one-brane model with the broken the 4D Poincar\a'{e} invariance necessarily contains
singular points\footnote{In the paper \cite{Cline:2001yt} it has been proved the no-go theorem which states that it is impossible to shield the singularity (where $B(\xi)=0$) from the brane by a horizon (where $A(\xi)=0$), unless the positive energy condition is violated in the bulk or on the brane. This statement is based on Eq. (7) of the paper \cite{Cline:2001yt} in the case of the $\mathbb{Z}_2$ symmetry (although, this symmetry condition is not a crucial point). Our solution with the restored $\mathbb{Z}_2$ symmetry (see the paragraph after our Eq. \rf{3.5}) coincides with the first example of the paper \cite{Cline:2001yt} (where in Eq. (8) we should keep only the potential of the scalar field playing the role of the bulk cosmological constant). Similarly to their result, we obtained $A(\xi)=-B(\xi)$, and the only horizon is possible at $\xi\to \infty$ where both $A(\xi)$ and $B(\xi)$ tend to zero. The authors of \cite{Cline:2001yt} demonstrated that the no-go theorem can be evaded if three-brane has a positive spatial curvature. We choose another way to solve this problem and introduce a second brane which cuts off all singular points.}. It is worth noting that the coordinate singularities, which deserve a special consideration as a separate paper, can be removed with a proper coordinate transformations. As it follows from our consideration below, these singularities appear only in a particular narrow range of parameters. Therefore, the Poincar\a'{e} invariance results in the curvature singularity except for a very narrow range of parameters. In our work, to avoid all singularities (physical as well as coordinate), we construct a regular solution by introducing a second brane in such a way that all singular points (both $B(\xi)=0$ and $A(\xi)=0$) do not lie between the branes. Obviously, this will be a generalization of the Randall-Sundrum solution (RSI) with two branes \cite{Randall:1999ee}.

\subsection{Two-brane model}
\label{subsec:B}

We assume that there is one more brane in addition to the brane at the point $\xi=0$. In general, we do not require the ${\mathbb Z}_2$ symmetry with respect to
the brane at $\xi =0$ and allow the fifth coordinate $\xi$ to run from $-L$ to $R$: $\xi\in[-L, R]$, with $L>0, R>0$ and $L\neq R$. The points $\xi =-L$ and
$\xi=R$ are identified with each other (see Fig.~\ref{plot1}), thus, $\xi$ parameterizes a topological torus $S^1$. We also assume that the bulk cosmological
constant has its own values in each sector of the bulk: $\Lambda_L$ and $\Lambda_R$ with $\Lambda_L, \Lambda_R <0$. Obviously, the  ${\mathbb Z}_2$ symmetry will
be restored for $\Lambda_L = \Lambda_R $. Then, according to Eqs. \rf{3.4} and \rf{3.5}, the metric coefficients in such a model can be written as follows:
%%%%%%%%%
\begin{figure}
  \includegraphics[width=0.45\textwidth]{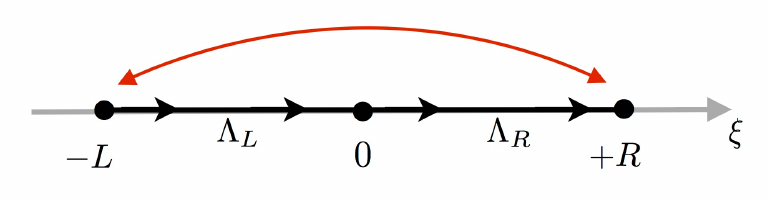}\\
  \caption{The schematic plot of the two-brane model. The red/upper left-right arrow indicates the points of identification. The black arrows show directions of
  normal vectors to the branes.}\label{plot1}
\end{figure}
%%%%%%
\ba{3.8}
B_L(\xi)&=&\beta_1e^{\mu\xi}+\beta_2e^{-\mu\xi},\,\, \mu^2\equiv-\cfrac{2}{3}\,\Lambda_L>0,\nn \\
A_L(\xi)&=&\alpha_1\,\cfrac{[B_L'(\xi)]^2}{B_L(\xi)}\,, \quad \xi\in[-L,0],
\ea
%%%%%%
and
%%%%%%%
\ba{3.9}
B_R(\xi)&=&b_1e^{m\xi}+b_2e^{-m\xi},\,\, m^2\equiv-\cfrac{2}{3}\,\Lambda_R>0,\nn \\
A_R(\xi)&=&a_1\,\cfrac{[B_R'(\xi)]^2}{B_R(\xi)}, \quad \xi\in[0,R].
\ea
%%%%%%
As we have mentioned above, in general, $\Lambda_L\neq\Lambda_R$, and, therefore, $\mu\neq m$.

These expressions are parameterized by four dimensional parameters ($\mu, L, m, R$) and six dimensionless parameters ($\alpha_1, \beta_{1,2}, a_1, b_{1,2}$). On
the other hand, the actual number of parameters in the model can be reduced. Indeed, the metric tensor is supposed to be well-defined and continuous at $\xi=0$
and at $(\xi=-L) \leftrightarrow (\xi=R)$:
%%%%%
\ba{3.10}
&{}&B_L(0)=B_R(0), \quad A_L(0)=A_R(0)\, ,\\
\label{3.11}
&{}&B_L(-L)=B_R(R), \quad A_L(-L)=A_R(R)\, .
\ea
%%%%%%
Without loss of generality we can also demand that
%%%%%%
\be{3.12}
B_L(0)=B_R(0)=-1,\quad A_L(0)=A_R(0)=1\, .
\ee
%%%%%%
Then, using these conditions \rf{3.12}, we obtain
%%%%%%
\ba{3.13}
\hspace*{-1cm}\beta_1&=&-(1+\beta_2)\, ,\qquad b_1=-(1+b_2), \\
\label{3.14}
\hspace*{-1cm}\alpha_1&=&-\mu^{-2}(1+2\beta_2)^{-2},\quad  a_1=-m^{-2}(1+2b_2)^{-2}.
\ea
%%%%%%%
With the help of Eqs. \rf{3.13}, it can be easily seen that the one-brane solution always has the singular points.
For example, $B_L(\xi)$, given by \rf{3.8}, always has zero for the range of parameters $ \beta_2 <-1$ and $\beta_2>0$. For $\beta_2\in (-1,0)$, zero of $B_L(\xi)$ is absent.
However, exactly for this interval of $\beta_2$ (i.e. for $\beta_2\in (-1,0)$), the metric coefficient $A_L(\xi)$ always has zero. Similar situation takes place for $B_R(\xi)$ and $A_R(\xi)$.

The condition $B_L(-L)=B_R(R)$ gives:
\ba{3.15}
\hspace*{-0.5cm}&-&(1+\beta_2)e^{-\mu L}+\beta_2e^{\mu L}=-(1+b_2)e^{mR}+b_2e^{-mR},\nn \\
\hspace*{-0.5cm}&{}& \ea
leading to
\ba{3.16} \label{3.16}
\hspace*{-0.5cm}&{}&\beta_2=[-(1+b_2)e^{mR}+b_2e^{-mR}+e^{-\mu L}](e^{\mu L}-e^{-\mu L})^{-1}.\nn \\
\hspace*{-0.5cm}&{}& \ea
From the equation $A_L(-L)=A_R(R)$ after cumbersome calculations we obtain:
%%%%%%
\ba{3.17}
\begin{aligned}
4&(e^{m R}-e^{-m R})(e^{-\mu L}-e^{-mR}-e^{mR}+e^{\mu L}) \\
\times&\left[  e^{mR}+2b_2 e^{mR}+b_2^2(e^{mR}-e^{-m R})\right] \\
\times& \left[e^{m R}+b_2(e^{-\mu L}+e^{\mu L}+2e^{m R})\right. \\
&\quad{}+\left.b_2^2(e^{-\mu L}+e^{-m R}+e^{m R}+e^{\mu L})\right]\\
\times&\left[b_2(e^{m R}-e^{-m R})+e^{m R}\right]^{-1} \\
\times&\left\{e^{\mu L}+e^{-\mu L}-2\left[b_2(e^{m R}-e^{-m R})+e^{m R}\right]\right\}^{-2} \\
\times& (1+2b_2)^{-2}=0\, .
\end{aligned}
\ea
%%%%%%%%
This equation is well-defined for all values of $b_2$, except for the following ones:
%%%%%%
\ba{3.18}
\hspace*{-0.5cm}&{}&b_2=-\cfrac{1}{2}, \quad b_2=\cfrac{1}{e^{-2m R}-1}, \quad b_2=\cfrac{e^{\mu L}+e^{-\mu L}-2e^{m R}}{2(e^{m R}-e^{-m R})}\, .\nn\\
\hspace*{-0.5cm}&{}&
\ea
%%%%%%%
Taking into account that $b_2, \mu, m, L, R$ are all real-valued, Eq. \rf{3.17} is satisfied if at least one of the following equations is satisfied:
%%%%%%%
\be{3.19}
e^{-\mu L}-e^{-mR}-e^{mR}+e^{\mu L}=0\, ,
\ee
%%%%%%%%
%%%%%%%%
\be{3.20}
e^{mR}+2b_2 e^{mR}+b_2^2(e^{mR}-e^{-m R})=0\, ,
\ee
%%%%%%%
%%%%%%%
\ba{3.21}
&{}&e^{m R}+b_2(e^{-\mu L}+e^{\mu L}+2e^{m R})\nn \\
&+&b_2^2(e^{-\mu L}+e^{-m R}+e^{m R}+e^{\mu L})=0\, .
\ea
%%%%%%%
In what follows we investigate these equations; namely, the first equation \eqref{3.19} in a subsection and then equations \eqref{3.20} and \eqref{3.21} in a
separate subsection.

\subsubsection{Equation \rf{3.19}}

It can be easily seen that Eq. \rf{3.19} is satisfied only if
%%%%%%%%
\be{3.22}
mR=\mu L \quad \Rightarrow \quad R=\cfrac{\mu}{m}\, L\, .
\ee
%%%%%%%%
Obviously, in the case $\mu=m$ the $\mathbb Z_2$ symmetry is restored. We note that this relation \rf{3.22} reduces the number of free parameters.
%Hereafter, it is convenient to introduce a new notation: $b\equiv b_2$.
Then, from Eq. \rf{3.15} we obtain:
%%%%%%
\ba{3.23}
&{}&\beta_1=b_2, \quad \beta_2=b_1= -(1+b_2), \nn \\
&{}&\alpha_1=-\cfrac{1}{\mu^2(1+2b_2)^2}, \quad a_1=-\cfrac{1}{m^2(1+2b_2)^2}\, .
\ea
%%%%%%
Consequently, the metric coefficients read
%%%%%
\ba{3.24}
A_L(\xi)&=&\frac{[b_2e^{\mu \xi}+(1+b_2)e^{-\mu\xi}]^2}{(1+2b_2)^2
   \left[(1+b_2)e^{-\mu\xi}-b_2e^{\mu\xi}\right]},\nn\\
&{}& \quad\quad\quad\quad\quad\quad\quad\quad\quad \xi\in[-L, 0],\\
\label{3.25} A_R(\xi)&=&\frac{[(1+b_2)e^{m\xi}+b_2 e^{-m \xi}]^2}{(1+2b_2)^2
   \left[(1+b_2)e^{m\xi}-b_2e^{-m\xi}\right]},\nn \\
&{}& \quad\quad\quad\quad\quad\quad\quad\quad\quad \xi\in\left[0, \cfrac{\mu}{m}\, L\right], \ea
%%%%%%%
and
%%%%%%
\ba{3.26}
\hspace*{-0.5cm}B_L(\xi)&=&b_2 e^{\mu\xi}-(1+b_2)e^{-\mu\xi}, \quad \xi\in[-L, 0]\, ,\\
\label{3.27}
\hspace*{-0.5cm}B_R(\xi)&=&-(1+b_2)e^{m\xi}+b_2 e^{-m\xi}, \quad \xi\in\left[0, \cfrac{\mu}{m}\, L\right].
\ea
%%%%%%%
Eqs. \rf{3.24}-\rf{3.27} clearly show that two values of the parameter $b_2$: $b_2=0,-1$ are special. For these values of $b_2$, the metric coefficients
$A_{L}(\xi)=-B_{L}(\xi)$ and  $A_{R}(\xi)=-B_{R}(\xi)$, and the 4D Poincar\a'{e} invariance is restored. This solution is well defined over the entire interval
$\xi\in [-L,R]$ (see Fig. 2) and, in the case $\mu=m$, is reduced to the RSI solution \cite{Randall:1999ee}. However, for other values of $b_2$ solutions can be
singular in some points of this interval. Now, we will define the allowed values of $b_2$ for which both $A_{L, R}$ and $B_{L, R}$ are not equal to zero at any
point of $\xi\in [-L, R]$.

As it follows from Eq. \rf{3.26}, $B_L(\xi)$ is equal to zero at
%%%%%%%%
\be{3.28}
\xi_{BL}=\cfrac{1}{2\mu}\,\ln\left(\frac{1 + b_2}{b_2}\right), \quad b_2\in(-\infty, -1)\cup(0, +\infty).
\ee
%%%%%%%
We require that the function $B_L(\xi)$ is nonzero in the interval $[-L, 0]$. This is possible either if $b_2\in[-1, 0]$, or if $\xi_{BL}>0$, or if $\xi_{BL}<-L$.
The condition $\xi_{BL}>0$ is equivalent to
%%%%%
\be{3.29}
\ln\left(\frac{1 + b_2}{b_2}\right)>0\quad\Leftrightarrow\quad b_2\in(0, +\infty),
\ee
%%%%%%
while the condition $\xi_{BL}<-L$ results in the condition
%%%%%%
\be{3.30}
\ln\left(\frac{1 + b_2}{b_2}\right)<-2\mu L \quad \Leftrightarrow \quad b_2\in \left(\cfrac{1}{e^{-2 \mu L}-1}, -1\right),
\ee
%%%%%%
where we have taken into account the condition that $\mu L >0$. Hence, in order for $B_L(\xi)$ to be nonzero at $[-L, 0]$, the parameter $b_2$ must belong to the
interval $I_1$:
%%%%%%
\be{3.31}
b_2\in I_1=\left(\cfrac{1}{e^{-2 \mu L}-1}, +\infty\right)\, .
\ee
%%%%%%%%

Following the similar steps by taking the condition $mR=\mu L$ into account in this case, we can demonstrate that $B_R(\xi)$ is not equal to zero at $[0, R]$, if
the parameter $b_2$ also lies within the interval $I_1$.

A similar analysis shows that the metric coefficients $A_L(\xi)$ and $A_R(\xi)$ are nonzero between branes (e.g., for $\xi\in [-L,R]$) if the parameter $b_2$
belongs to the interval $I_2$:
%%%%%%%
\be{3.32}
b_2\in I_2=\left(-\infty, -\cfrac{1}{1+e^{-2\mu L}}\right)\cup(-1/2, +\infty)\, .
\ee
%%%%%%%%
Here, we also excluded the value $b_2=-1/2$ at which $A_{L,R}(\xi)$ becomes singular.

Finally, for all metric coefficients in our model to be non-singular and nonzero, the parameter $b_2$ must belong to the interval $I$:
%%%%%%
\ba{3.33} &{}&b_2\in I=I_1\cap I_2\nn\\
&=&\left(\cfrac{1}{e^{-2\mu L}-1}, -\cfrac{1}{e^{-2\mu L}+1}\right)\cup (-1/2, +\infty). \ea
%%%%%%
In what follows, $b_2$ is assumed to belong to this range of parameters. It's also worth noting here that none of the prohibited values \rf{3.18} lies within this
interval.

Therefore, we have constructed a class of non-singular solutions \rf{3.24}-\rf{3.27} that are well-defined, continuous and nonzero over the whole domain $S^1$
parameterized by $\xi\in[-L, (\mu/m)L]$. The free parameters $\mu, m, L$ are strictly positive and $b_2$ can take any values from the set $I$ given in \rf{3.33}.
% An example of such solution is drawn in Fig. 3.

The metric coefficients \rf{3.24}-\rf{3.27} are continuous at $\xi=0$ and $(\xi=-L) \leftrightarrow (\xi=R)$. However, their derivatives are not. The ``jumps'' of
the derivatives are interpreted as the presence of branes filled with some matter content. The energy-momentum tensor (EMT) of the matter on the branes is defined
via Israel junction conditions:
%%%%%
\ba{3.34}
&{}&\bigl.\left(K_{mn}-g_{mn} K\right)\bigr]^{\xi=+0}_{\xi=-0}=\kappa \tau_{mn}(\xi=0)\, ,\nn\\
&{}&\bigl.\left(K_{mn}-g_{mn} K\right)\bigr]^{\xi=-L+0}_{\xi=R-0}=\kappa \tau_{mn}(\xi=-L)\, ,
\ea
%%%%%%
where the extrinsic curvature tensor in the chosen coordinates reads $K_{mn}=-(1/2) g_{mn}'$. The tensor $\tau_{mn}$ is interpreted as the EMT of matter localized
on two branes (``1'' at $\xi=0$ and ``2'' at $\xi=-L$). We assume that each brane is filled with a perfect fluid. Therefore:
%%%%%%
\ba{3.35}
\hspace*{-1cm}&{}&\tau_{00}(\xi=0,-L)=\epsilon_{(1,2)} A(\xi=0,-L)\, ,\\
\label{3.36} \hspace*{-1cm}&{}&\tau_{ii}(\xi=0,-L)=-\pi_{(1,2)} B(\xi=0,-L), \  i=1, 2, 3, \ea
%%%%%%%
where the quantities $\epsilon_{(k)}$ and $\pi_{(k)}$, $k=1, 2$, are interpreted as energy density and pressure of the fluids on the branes ``1'' and ``2'',
respectively. We also introduce an EoS parameter for each fluid:
%%%%%%
\be{3.37}
\pi_{(k)}=\Omega_{(k)}\epsilon_{(k)},\quad k=1, 2\, .
\ee
%%%%%%%
For the metric coefficients \rf{3.24}-\rf{3.27} we obtain
%%%%%%%
\ba{3.38}
\epsilon_{(1)}&=&\cfrac{3}{2\kappa} \,(1 + 2 b_2) (m + \mu)\, ,\\
\label{3.39} \Omega_{(1)}&=&-1 +\cfrac{8}{3} \frac{b_2(1+b_2)}{(1 + 2 b_2)^2}\, , \ea
%%%%%%
and
%%%%%
\ba{3.40}
\epsilon_{(2)}&=&\cfrac{3}{2\kappa}\,\cfrac{b_2e^{-\mu L}+(1+b_2)e^{\mu L}}{b_2e^{-\mu L}-(1+b_2)e^{\mu L}}\,(m+\mu)\, ,\\
\label{3.41}
\Omega_{(2)}&=&-1-\frac{8}{3}\cfrac{b_2(1+b_2)}{[b_2 e^{-\mu L}+(1+b_2)e^{\mu L}]^2}\, .
\ea
%%%%%%%
Clearly, the quantities $\epsilon_{(k)}$, $\pi_{(k)}$, $\Omega_{(k)}$, $k=1, 2$, are well-defined and nonzero for all the allowed values of the free parameters.

It is noteworthy that although $\epsilon_{(1)}$ and $\Omega_{(1)}$ that define the matter on the brane ``1''  do not depend on $L$, the corresponding quantities
$\epsilon_{(2)}$ and $\Omega_{(2)}$ on the brane ``2'' do. This dependence means that, generally, not only the value of the energy density on the second brane is
fine tuned to the distances between branes, but also the EoS parameter of this matter depends on $L$ (or, taking into account the relation $\mu L=mR$, on $R$).
Obviously, in the particular cases of the restored 4D Poincar\a'{e} invariance $b_2=0,-1$, we reproduce the results of the RSI model. Here, we have the vacuum EoS
on both branes: $\Omega_{(1)}=\Omega_{(2)}=-1$. It is well known that RSI model requires one of the branes to be filled with negative energy density. This
situation holds also for the general case of the broken Poincar\a'{e} invariance. Here, for the allowed set of parameters:
$\text{sign}(\epsilon_{(1)})=-\text{sign}(\epsilon_{(2)})$. To demonstrate it, we can mention that according to Eq. \rf{3.38}
%%%%%%
\ba{3.42}
&{}&\epsilon_{(1)}>0\quad  \Leftrightarrow \quad b_2 >-1/2 \quad \Leftrightarrow b_2\in I_3, \\
\label{3.43}
&{}&\epsilon_{(1)}<0\quad \Leftrightarrow \quad b_2<-1/2 \quad \Leftrightarrow b_2\in I_4,
\ea
%%%%%
where
%%%%%%
\be{3.44} I_3 = \left(-\cfrac{1}{2}, +\infty\right),\, \ I_4= \left(\cfrac{1}{e^{-2\mu L}-1}, -\cfrac{1}{e^{-2\mu L}+1}\right). \ee
%%%%%%
To define the interval $I_4$ we took into account that the parameter $b_2$ should belong to the allowed regions \rf{3.33}. Then, $I_4 = I \setminus (-1/2,
+\infty)$. The similar analysis of Eq. \rf{3.40} demonstrates that
%%%%%%
\ba{3.45}
&{}&\epsilon_{(2)}<0\quad \Leftrightarrow \quad b_2\in I_3, \\
\label{3.46}
&{}&\epsilon_{(2)}>0 \quad \Leftrightarrow \quad b_2\in I_4\, .
\ea
%%%%%
Hence, $\epsilon_{(1)}$ and $\epsilon_{(2)}$ always have opposite signs{\footnote{\label{footSMS}In the paper \cite{Shiromizu:1999wj} it was shown that in the projective
approach the sign of the effective four-dimensional gravitational constant is defined by the sign of the vacuum energy density in the brane. To arrive at this
conclusion, the authors supposed that there is the vacuum energy in the brane with the EoS parameter $\Omega=-1$. So, the Newton's gravitational constant has the
wrong sign if the vacuum energy density is negative. In our model, the energy densities of matter in the branes are defined by $\epsilon_{1,2}$. We did not
postulate the form of the matter in the branes but defined it from the Israel junction condition and found that, similarly to the RSI model, one of the branes
always has negative sign of the energy density. However, the EoS parameters differ from the vacuum-{like value $-1$ (see Eqs. \rf{3.39}, \rf{3.41}).
Therefore, we cannot apply directly the results of the paper \cite{Shiromizu:1999wj}. To conclude about the form of gravitational interaction in the branes, we need to study
the linearized perturbations of the considered model \cite{Rubakov:2001kp}. However, this investigation is out of the scope of the present paper.}}. It is worth noting
that the null-energy condition (NEC) $\epsilon_{(k)} + \pi_{(k)}=\epsilon_{(k)}\left[1+\Omega_{(k)}\right]\geq0,\; k=1, 2$, can be fulfilled for both branes. To
demonstrate it, we consider first the brane ``1'' with the energy density and the EoS parameter given by Eqs. \rf{3.38} and \rf{3.39}. In this case the NEC reads
%%%%%%%
\be{3.47d}
\epsilon_{(1)}\left[1+\Omega_{(1)}\right]\geq0 \quad \Rightarrow \quad \frac{b_2(1+b_2)}{1 + 2 b_2}\geq0\, .
\ee
%%%%%%
This inequality is satisfied for  $b_2\in [-1, -1/2)\cup [0, +\infty)$. Now, let us turn to the brane ``2'' with the energy density and the EoS parameter given by
Eqs. \rf{3.40} and \rf{3.41}. Here, the NEC is reduced to the following inequality:
%%%%%%%
\be{3.48d}
\cfrac{b_2(1+b_2)}{\left[b_2(e^{-\mu L}-e^{\mu L})-e^{\mu L}\right]\left[b_2(e^{-\mu L}+e^{\mu L})+e^{\mu L}\right]}\leq0\, ,
\ee
%%%%%%
which admits the solution $b_2\in \left(-\infty, \cfrac{1}{ e^{-2\mu L}-1}\right)\cup \left[-1, -\cfrac{1}{ e^{-2\mu L}+1}\right)\cup \left[0, +\infty\right)$.

Taking into account the interval \rf{3.33} for the allowed values of $b_2$, we see that both branes ``1'' and ``2'' satisfy the NEC for the values
%%%%%%
\be{3.49d}
b_2\in \left[-1, -\cfrac{1}{ e^{-2\mu L}+1}\right)\cup \left[0, +\infty\right)\, .
\ee
%%%%%%%%

In Fig.~\ref{plot2} and Fig.~\ref{plot3}, we present examples of the metric coefficients $A(\xi)$ and $B(\xi)$ in the case of restored (Fig.~\ref{plot2}) and
broken (Fig.~\ref{plot3}) 4D Poincar\a'{e} invariance. Here, the values of the parameter $b_2$ are taken  from the interval $I$ \rf{3.33}.

%%%%%%%%%%%%%%%%%%%%%%%%%%%%
\begin{figure*}[htbp]
  \hfil\includegraphics[width=2.7 in]{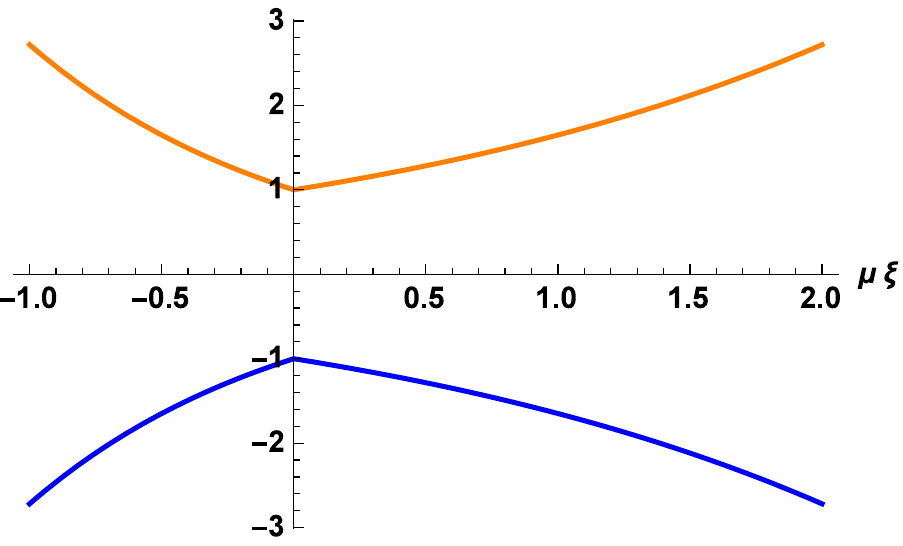}\hfil
  \hfil\includegraphics[width=2.7 in]{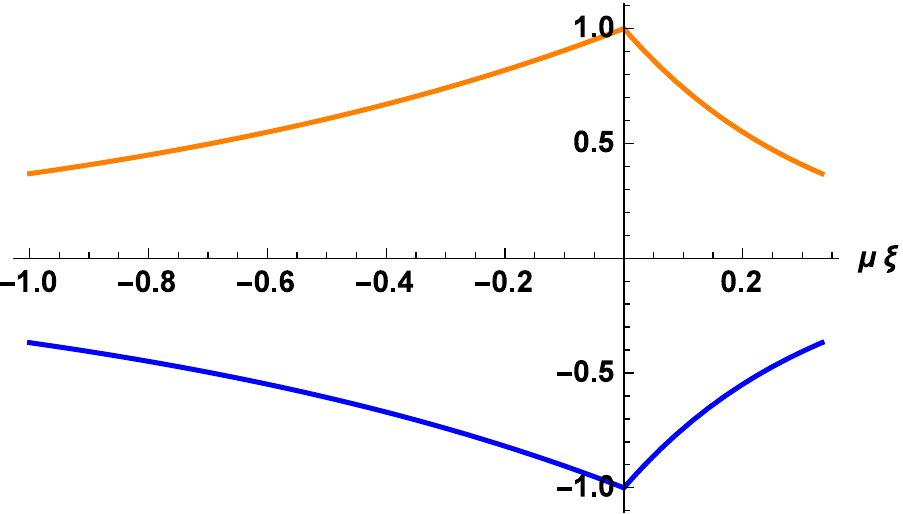}\hfil
  \caption{
            \textbf{Models with restored 4D Poincar\a'{e} invariance.} Orange/top and blue/bottom lines represent $A(\xi)$ and $B(\xi)$ metric coefficients,
            respectively. \textsf{Left graph}: $\mu L=1$, $\mu/m=2$, $b_2=0$:
            $(\kappa/\mu)\epsilon_{(1)}=-(\kappa/\mu)\epsilon_{(2)}=2.25,\, \Omega_{(1)}=\Omega_{(2)}=-1.$ \textsf{Right graph}: $\mu L=1$, $\mu/m=1/3$, $b_2=-1$:
            $(\kappa/\mu)\epsilon_{(1)}=-(\kappa/\mu)\epsilon_{(2)}=-6,\, \Omega_{(1)}=\Omega_{(2)}=-1.$
  }\label{plot2}
\end{figure*}
%%%%%%%%%%%%%%%%%%%%%%%%%%%%%
\begin{figure*}[htbp]
  \hfil\includegraphics[width=2.7 in]{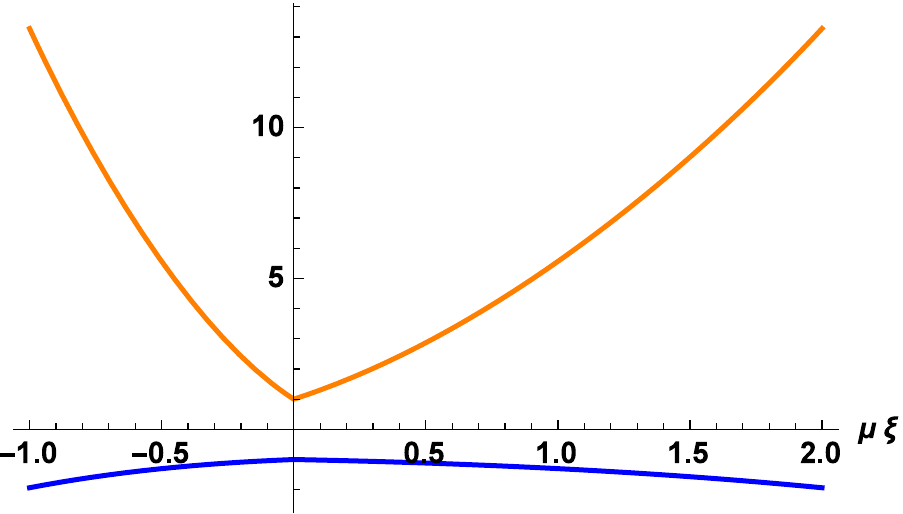}\hfil
  \hfil\includegraphics[width=2.7 in]{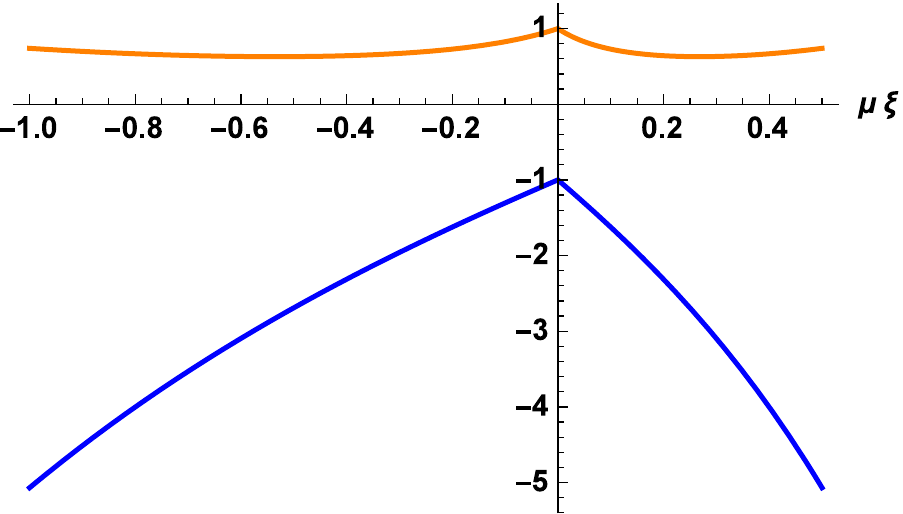}\hfil
  \caption{\textbf{Models with broken 4D Poincar\a'{e} invariance.} Orange/top and blue/bottom lines represent $A(\xi)$ and $B(\xi)$ metric coefficients,
  respectively.  \textsf{Left graph}: $\mu L=1$, $\mu/m=2$ and $b_2=-1/3$:
 $(\kappa/\mu)\epsilon_{(1)}=0.75,\,  (\kappa/\mu)\epsilon_{(2)}\approx-1.96,\, \Omega_{(1)}\approx-6.33, \Omega_{(2)}\approx-1.21.$
 \textsf{Right graph}: $\mu L=1$, $\mu/m=1/2$ and $b_2=1$:
 $(\kappa/\mu)\epsilon_{(1)}=13.5,\, (\kappa/\mu)\epsilon_{(2)}\approx-5.15,\, \Omega_{(1)}\approx-0.41, \Omega_{(2)}\approx-0.84.$}\label{plot3}
\end{figure*}
%%%%%%%%%%%%%%%%%%%%%%%%%%%%

\subsubsection{Equations \rf{3.20} and \rf{3.21}}

To conclude this section, we consider Eqs. \rf{3.20} and \rf{3.21} and briefly show that there are no any new non-singular solutions for the metric coefficients
in these cases. We start from Eq. \rf{3.20} which we consider as a quadratic equation with respect to $b_2$.
%(in this subsection we came back again to the original notation $b_2$ instead of $b$).
Then, this equation has the following solutions:
%%%%%%%%
\be{3.47}
b_2=\cfrac{e^{m R}}{1 - e^{m R}}\,, \quad b_2=-\cfrac{e^{m R}}{1 + e^{m R}}\,.
\ee
%%%%%%%
First, it can be easily seen that if we consider $\mu L= m R$, as it took place in the previous subsection, then these values of $b_2$ do not belong to the
allowed interval $I$ given in \rf{3.33}. So, we consider the case $\mu L\neq m R$. Taking into account Eqs. \rf{3.13} and \rf{3.15}, we obtain for the first root
in \rf{3.47}:
%%%%%%%%
\be{3.48}
\beta_2=\cfrac{1}{e^{\mu L}-1}\, ,\quad \beta_1=-\cfrac{e^{\mu L}}{e^{\mu L}-1}\,.
\ee
%%%%%%%%
Then
%%%%%%%
\be{3.49}
B_L(\xi)=\beta_1e^{\mu \xi}+\beta_2 e^{-\mu \xi}=\cfrac{e^{-\mu\xi}-e^{\mu(\xi+L)}}{e^{\mu L}-1}\,, \quad \xi\in[-L, 0].
\ee
%%%%%%%
This function equals zero at the point $\xi=-L/2$, irrespectively of the choice of the value of $L$.
%If we restore the mirror symmetry, then $B$or strictly medially between the branes, if $m=\mu$!).
Hence, for this choice of $b_2$ regular solutions do not exist.

If we consider the second root for $b_2$ in Eq. \rf{3.47}, we get for $\beta_2$ and $\beta_1$:
%%%%%%%
\be{3.50}
\beta_2=-\cfrac{1}{e^{\mu L}+1}\,,\quad
\beta_1=-\cfrac{e^{\mu L}}{e^{\mu L}+1}\,,
\ee
%%%%%%%%
and, hence, the functions $B_L(\xi)$ and $A_L(\xi)$ read
%%%%%%
\ba{3.51}
B_L(\xi)&=&-\cfrac{e^{-\mu \xi}+e^{\mu(\xi+ L)}}{e^{\mu L}+1}\, ,\\
\label{3.52}
A_L(\xi)&=&-\alpha_1 \cfrac{\mu^2}{e^{\mu L}+1}\frac{\left[e^{-\mu\xi}-e^{\mu (\xi+L)}\right]^2}{e^{-\mu \xi}+e^{\mu(\xi+L)}}\, ,
\ea
%%%%%%%%
where $\xi\in[-L, 0]$. These equations show that the function $B_L(\xi)$ does not go to zero anywhere in this interval. However, $A_L(\xi)$ is equal to zero at
$\xi=-L/2$. Thus, both of the solutions of Eq. \rf{3.20} result in zero metric coefficients in the bulk between the branes.

Let us now turn our attention to Eq. \rf{3.21}. It also has two solutions:
%%%%%%%
\be{3.53}
b_2=-\frac{e^{m R}}{e^{\mu
   L}+e^{m R}}, \quad b_2= -\frac{e^{\mu  L+m R}}{e^{\mu
   L+m R}+1}\, .
\ee
%%%%%%%

It is not difficult to check that if we set $\mu L = mR$, these values of $b_2$ again belong to the prohibited region: $b_2\not\in I$, where the interval $I$ is
given by \rf{3.33}. Therefore, we will again assume that the free parameters $m, \mu, L, R$ do not obey the fine tuning condition $\mu L= m R$, and try to
determine whether any regular geometries exist for the roots \rf{3.53}.

In the case of the first root in \rf{3.53}, Eqs. \rf{3.16} and \rf{3.13} give
%%%%%%%
\be{3.54}
\beta_2=-\cfrac{e^{m R}}{e^{\mu L}+e^{m R}} =b_2,\quad
b_1=-\cfrac{e^{\mu L}}{e^{\mu L}+e^{m R}}=\beta_1.
\ee
%%%%%%%%
It is clear that $\text{sign}(\beta_1)=\text{sign}(\beta_2)$ and $\text{sign}(b_1)=\text{sign}(b_2)$ and therefore $B_L(\xi)$ and $B_R(\xi)$ are nowhere zero. Due
to the same reason $A_L(\xi)$ and $A_R(\xi)$ can be zero:
%%%%%%
\ba{3.55}
A_L(\xi)&\sim& [-e^{\mu L}e^{\mu \xi}+e^{m R}e^{-\mu \xi}]^2, \quad \xi\in[-L, 0],\\
\label{3.56}
A_R(\xi)&\sim& [-e^{\mu L}e^{m \xi}+e^{m R}e^{-m \xi}]^2, \quad \xi\in[0, R].
\ea
%%%%%%%
$A_L(\xi)$ is zero at $\xi_L=(mR-\mu L)/(2\mu)$, while $A_R(\xi)$ is zero at $\xi_R=(mR-\mu L)/(2m)$.

$A_L(\xi)$ may be nonzero over $[-L, 0]$ if $\xi_L>0$ or $\xi_L<-L$. The first inequality is equivalent to $R>(\mu/m) L$, while the second one is equivalent to
$R<-(\mu/m)L$, and, indeed, is not valid for any reasonable values of the free parameters.

$A_R(\xi)$ may be nonzero over $[0, R]$ if $\xi_R<0$ or $\xi_L>R$. The first inequality is equivalent to $R<(\mu/m) L$, while the second one is again equivalent
to $R<-(\mu/m)L$ (with empty solution set).

Therefore, the only possibility for $A(\xi)$ to be nonzero everywhere over $[-L, R]$ is when $\xi_L>0$ and $\xi_R<0$ simultaneously. However, this requires $R$ to
satisfy both inequalities $R>(\mu/m) L$ and $R<(\mu/m) L$, which is impossible. Hence, the first root of $b_2$ in \rf{3.53} is unsatisfactory. Similar analysis
for the second root results in the same conclusion.

Hence, the only possible regular, continuous and nowhere zero solutions for the metric coefficients are given by Eqs. \rf{3.24}-\rf{3.27} where parameter $b$
belongs to the interval $I$ \rf{3.33}.

%%%%%%%%%%%%%%%%%%%%%%%%%%%%%%%%%%%%%%%%%%%%%%%%%%%%%%%%%%%%%%%%%%%%%%%%%%%%%%%%%%%
\section{Bulk with perfect fluid}
\label{sec:bulk}

Now, we assume that bulk is filled with the perfect fluid with EMT of the form \rf{2.2}. This EMT can be also written as
%%%%%%
\ba{4.1}
&{}&T^M_N=\ve\left(\delta^M_0\delta_N^0-\om_0\sum_{\td\mu=1}^3\delta^M_{\td\mu}\delta^{\td\mu}_N-
\om_1\delta^M_4\delta^4_N\right),\nn\\
&{}&M,N=0,1,2,3,4 \, . \ea
%%%%%
Then the conservation equation $\nabla_M T^M_N=0$ is reduced to a system of equations
%%%%%%%
\ba{4.2}
\hspace*{-1cm}&{}&\de_0\ve=0\, ,\\
\label{4.3}
\hspace*{-1cm}&{}&\om_0\de_{\td\nu}\ve=0,\quad \td\nu =1,2,3\, ,\\
\label{4.4}
\hspace*{-1cm}&{}&\om_1\ve'+\om_1\ve\,\left[\cfrac{A'}{2A}\,+\cfrac{3}{2}\,\cfrac{B'}{B}\right]
+\ve\left[\cfrac{A'}{2A}-\om_0\,\cfrac{3}{2}\,\cfrac{B'}{B}\right]=0\, ,
\ea
%%%%%%%
where we took into account our metric ansatz \rf{2.1}. From \rf{4.2} we find that $\ve$ must be static. If $\om_0\neq 0$, then \rf{4.3} results in a conclusion
that the energy density may depend only on $\xi$: $\ve=\ve(\xi)$. However, it is not necessary to suppose that $\om_0\neq 0$ to arrive at this conclusion. The
similar result follows from Eq. \rf{2.3} for our metric ansatz \rf{2.1}.

It is convenient to introduce new functions $b(\xi)\equiv B'/B$ and $a(\xi)\equiv A'/A$. Then, Eqs. \rf{2.3}-\rf{2.5} and \rf{4.4} take the form, respectively:
%%%%%%%
\ba{4.5}
\hspace*{-1cm}&{}&b'+b^2=-\cfrac{2}{3}\,(\La+\ve), \\
\label{4.6}
\hspace*{-1cm}&{}&-(b'+b^2)-\cfrac{1}{2}\,(a'+a^2)+\cfrac{1}{4}\,\left(b-a\right)^2 =\La-\om_0\ve , \\
\label{4.7}
\hspace*{-1cm}&{}&b(b+a)=-\cfrac{4}{3}\,(\La-\om_1\ve), \\
\label{4.8}
\hspace*{-1cm}&{}&\om_1\ve'+
\ve\left[\cfrac{1}{2}\,(1+\om_1)a+\cfrac{3}{2}\,\,(\om_1-\om_0)b\right]=0.
\ea
%%%%%%
From \rf{4.5} we get:
%%%%%%
\be{4.9}
\ve=-\cfrac{3}{2}\,(b'+b^2)-\La, \quad \ve'=-\cfrac{3}{2}\,(b''+2bb')\, .
\ee
%%%%%%

\

\subsection{$\om_1\neq -1,0$}
\label{subsec:4A}
\

Simple analysis of Eqs. \rf{4.7}-\rf{4.9} demonstrates that the function $b$ satisfies the following equation:
%%%%%%
\ba{4.10}
\hspace*{-1cm}&{}&\cfrac{2\om_1}{1+\om_1}\,
\cfrac{b''+2bb'}{b'+b^2+(2/3)\La} \\
\hspace*{-0.5cm}&=&\cfrac{2}{b}\left[\cfrac{2}{3}\,\La (1+\om_1)+\om_1 b'\right]+\left[1+2\om_1+\cfrac{3(\om_0-\om_1)}{1+\om_1}\right]b,\nn
\ea
%%%%%%
which can be resolved with respect to $b''$:
%%%%%
\ba{4.11}
\hspace*{-1cm}&{}&b''=-\cfrac{1}{\om_1}\left(-\cfrac{1}{2}+\om_1-2\om_1^2-\cfrac{3}{2}\,\om_0\right)b'b\nn\\
&+&\cfrac{1}{\om_1}\left(\cfrac{1}{2}+\om_1^2
+\cfrac{3}{2}\,\om_0\right)b^3\nn\\
&+&\cfrac{2}{3\om_1}\,\La(1+3\om_1+2\om_1^2)\cfrac{b'}{b}+(1+\om_1)\left(\cfrac{b'}{b}\right)^2b\nn\\
&+&\cfrac{4}{9\om_1}\,\La^2(1+\om_1)^2\,\cfrac{1}{b}\nn\\
&+&\cfrac{b}{\om_1}\left[\cfrac{4}{3}\,\La\om_1(1+\om_1)+\La(1+\om_0)\right]. \ea
%%%%%
It is not difficult to show that the system of Eqs. \rf{4.6}, \rf{4.7} and \rf{4.9} results in the same equation for $b$. Therefore, two of equations of the
system \rf{4.5}-\rf{4.8} are equivalent.

Eq. \rf{4.11} is an autonomous second order equation of the form
%%%%%%%
\be{4.12}
b''=\alpha_1b'b+\alpha_2b^3
+\alpha_3\cfrac{b'}{b}
+\alpha_4\cfrac{(b')^2}{b}
+\alpha_5\cfrac{1}{b}+\alpha_6b\, ,
\ee
%%%%%%%
where the values of the constants $\alpha$ are obvious. Via introduction of a new variable $u(b)\equiv b' \Rightarrow b''=\dot u u$ (where the dots denote
derivatives with respect to $b$) we reduce its order:
%%%%%%
\be{4.13}
\dot u=
\cfrac{\alpha_4}{b}\,u +\left(\alpha_1b +\cfrac{\alpha_3}{b}\right)+\left(\alpha_2b^3+\alpha_6b+\alpha_5\cfrac{1}{b}\right)\frac{1}{u},
\ee
%%%%%%
which is an equation of the form $y'=f_1(x)y+f_0(x)+f_{-1}(x)y^{-1}$. Generally, it cannot be solved by quadrature. The special cases $f_0\equiv0$ and
$f_{-1}\equiv0$ reduce this equation to the Bernoulli equation and the linear equation, respectively.

In our case, the condition $f_0\equiv0$ corresponds to $\alpha_1=0, \alpha_3=0$, and we get the system:
%%%%%%%
\ba{4.14}
         \alpha_1&\sim&-\cfrac{1}{2}+\om_1-2\om_1^2-\cfrac{3}{2}\,\om_0=0, \nn \\
         \alpha_3&\sim&1+3\om_1+2\om_1^2=0.
\ea
%%%%%%%
These equations are compatible only if $\om_0=-1, \,\,\om_1=-\cfrac{1}{2}$. On the other hand, the condition $f_{-1}=0$ is equivalent to
the system:
%%%%%%%
\ba{4.15}
         \alpha_2&\sim&\cfrac{1}{2}+\om_1^2+\cfrac{3}{2}\,\om_0=0, \nn \\
         \alpha_5&\sim&1+\om_1=0,\\
         \alpha_6&\sim&\cfrac{4}{3}\,\La\om_1(1+\om_1)+\La(1+\om_0)=0\nn
         \ea
%%%%%%%
with the only solution $\om_0=\om_1=-1$, which is prohibited by our requirement $\om_1\neq-1$. Therefore, we can solve \rf{4.5}-\rf{4.8} by quadrature in the case
of anisotropic fluid $\om_0=-1, \om_1=-1/2$ (vacuum in 4D and tension along the extra coordinate). However, the obtained expression for $u(b)$ does not allow to
solve $b'(\xi)=u(b(\xi))$ analytically{\footnote{Quite similar situation takes place in the case $\om_1=-1$. As it can be easily seen from Eqs. \rf{4.6}, \rf{4.7}
and \rf{4.9}, here Eq. \rf{4.11} should be replaced with the following one: $b'' = - b \left[(1 + \omega_0) (2\Lambda_5 + 3 b^2) + (7 + 3 \omega_0) b'\right]/2$.
This equation is solvable, e.g., in the cases $\om_0=-1$ and $\om_0=-7/3$. The first case is trivially reduced to the RSII case with renormalized bulk
cosmological constant. In the second case, we can introduce a new variable $u(b)\equiv b'$ and solve the first order differential equation with respect to $u(b)$.
However, we cannot invert the equation  $b'(\xi)=u(b(\xi))$ and solve it analytically with respect to $B(\xi)$.}}. Therefore, Eq. \rf{4.12} (or, equivalently,
\rf{4.13}) should be considered as a master equation for numerical studies of the considered brane world models for arbitrary bulk perfect fluid EoS parameters
except $\om_1\neq -1,0$. Nevertheless, there are also analytical solutions for particular values of the EoS parameters which are of physical interest. Below, we
consider two such solutions.

\subsection{$\om_1=0,\, \om_0\neq-\frac{1}{3}$}
\label{subsec:4B}

Now, we consider the special case $\om_1=0$.  Eq. \rf{4.10} (which is valid for such value of $\om_1$) results in the following relation:
%%%%%%%
\be{4.16}
b^2=-\cfrac{4}{3}\,\cfrac{\La}{(1+3\om_0)}=\mathrm{const},
\ee
%%%%%%
which is physically meaningful for $\La\neq0$ and $\om_0\neq-1/3$. Then, the system of Eqs. \rf{4.5}-\rf{4.9} results in the following fine-tuning condition for
the energy density of the perfect fluid:
%%%%%%
\be{4.17} \ve=\La\,\cfrac{1-3\om_0}{1+3\om_0}\, ,\quad \om_0\neq -\frac{1}{3}\, , \ee
%%%%%%
and solutions for the metric coefficients:
%%%%%%
\ba{4.18}
B(\xi)&=&B_0 \exp\left(\pm\sqrt{-\cfrac{4}{3}\,\cfrac{\La}{(1+3\om_0)}}\,\,\xi\right), \nn\\
A(\xi)&=&A_0[B(\xi)]^{3\om_0}. \ea
%%%%%%%
These functions are real-valued only if $\La(1+3\om_0)<0$. Therefore, if $\La <0$, the energy density $\varepsilon$ can be both negative
$\varepsilon <0$ (it happens for $-1/3<\om_0<1/3$) and positive $\varepsilon >0$ (for $\om_0>1/3$). However, if $\La>0$, the energy density
can be only negative $\varepsilon<0$ (for $\om_0<-1/3$). It is worth noting that the case $\om_0=1/3$ coincides formally with the RSII solution.

To restore the Minkowski metric on the section $\xi=0$, we normalize solutions \rf{4.18} as follows: $A(0)=1$ and $B(0)=-1$ which immediately yields: $B_0=-1$ and
$A_0=(-1)^{-3\om_0}$. Let us choose solutions decaying at $\xi\to +\infty$. This corresponds to the  minus in the exponent \rf{4.18}. Then, $A(\xi)$ reads
%%%%%%%
\be{4.19}
A(\xi)= \exp\left(-3\omega_0\sqrt{-\cfrac{4}{3}\,\cfrac{\La}{(1+3\om_0)}}\,\,\xi\right)\, .
\ee
%%%%%%%
Obviously, the decaying solution will take place only for $\om_0>0$. In this case the bulk cosmological constant can be only negative $\La<0$. Since the scalar
curvature of this model is $R=2 \Lambda_5 (2 + 3 \omega_0 + 3 \omega_0^2)/(1 + 3 \omega_0)$, only the spaces with $R<0$ are described by the considered model.

Now we follow the standard procedure to construct a one-brane model. The bulk is taken to be infinite and parameterized by $\xi\in\mathbb R$. For the sake of
mathematical generality, we again break the mirror symmetry $\xi\mapsto -\xi$ via introduction of two bulk regions, $\xi>0$ and $\xi<0$, each being characterized
by its own set of free parameters $\Lambda, \omega_0$:
%%%%%%%
\ba{4.20}
\hspace*{-1cm}&{}&\om_R\equiv \om_{0,R}, \, \Lambda_R\equiv \Lambda_{5,R}, \, m \equiv \sqrt{-\cfrac{2}{3}\,\Lambda_R}>0,\; \xi>0,\\
\label{4.21}
\hspace*{-1cm}&{}&\om_L\equiv \om_{0,L}, \, \Lambda_L\equiv \Lambda_{5,L}, \, \mu \equiv \sqrt{-\cfrac{2}{3}\,\Lambda_L}>0,\; \xi<0.
\ea
%%%%%%%

\vspace*{0.1cm}

\noindent In general, we suppose that $\omega_L\neq\omega_R$, $\mu\neq m$. The $\mathbb Z_2$-symmetry of the bulk is restored only if these inequalities become
equalities.

The metric coefficients of two regions are continuously glued together along the Minkowski brane located at $\xi=0$. The matter content on the brane can be
determined from the Israel junction conditions similar to Eqs. \rf{3.34}-\rf{3.37}. Therefore, for the energy density and pressure/tension on the brane we get:
%%%%%%
\ba{4.22}
\epsilon &=&-\cfrac{3}{\sqrt{2}\, \kappa}\left(\cfrac{m}{\sqrt{1+3\omega_R}}+\cfrac{\mu}{\sqrt{1+3\omega_L}}\right), \\
\label{4.23} \pi &=& \cfrac{1}{\sqrt{2}\, \kappa}\,\left( \cfrac{2+3\omega_R}{\sqrt{1+3\omega_R}}\, m+\cfrac{2+3\omega_L}{\sqrt{1+3\omega_L}}\, \mu \right). \ea
%%%%%%
Since $\om_{L,R}>0$, both the energy density $\epsilon$ and the equation of state parameter $\Omega =\pi/\epsilon$ are negative. These equations show that the
matter on the brane is fine-tuned not only to the ``cosmological constants'' $\mu, m$ but also to the parameters of EoS of bulk matter $\omega_{L, R}$. In the
case of restored $\mathbb Z_2$-symmetry ($\mu=m$ and $\omega_L=\omega_R=\omega_0$) the expressions are simplified as follows:
%%%%%%
\ba{4.24}
\epsilon&=&-\cfrac{3 \sqrt{2} m}{\kappa\sqrt{1 + 3 \omega_0}}, \quad
\pi=\sqrt{2}m\cfrac{2+3\omega_0}{\kappa\sqrt{1+3\omega_0}}\, , \\
\label{4.26}
\Omega&=&-\cfrac{2}{3}-\omega_0\, .
\ea
%%%%%%
It can be easily seen that the NEC condition $\epsilon +\pi \geq 0$ is satisfied for $\omega_0\geq 1/3$. As we already mentioned above, the case
$\om_0=1/3$ formally coincides with the RSII solution.

The characteristic behaviour of the metric coefficients $A(\xi)$ and $B(\xi)$ in the case of restored $\mathbb Z_2$-symmetry is depicted in Fig.~\ref{plot4}.
\begin{figure}
  \includegraphics[width=3 in]{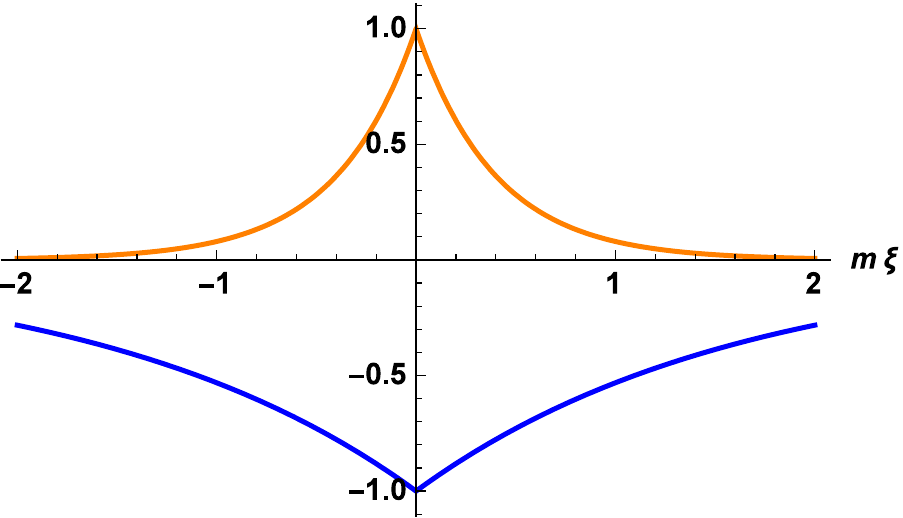}\\
  \caption{The plot of the metric coefficients $A(\xi)$ (orange/top line) and $B(\xi)$ (blue/bottom line) in the case $\omega_1=0$ and $\omega_0=4/3$. Here,
  $(\kappa/m)\epsilon=-3\sqrt{2/5}$ and $\Omega=-2$. 4D Poincar\a'{e} invariance is broken.}\label{plot4}
\end{figure}

\subsection{Thick brane: $a=b,\, \om_1\neq 0,-1,\, \om_0=-1$}
\label{subsec:4C}

Now, let us restore the 4D Poincar\a'{e} invariance: $a=b$. We also exclude the value $\om_1=0$ since, as it easily follows from Eqs. \rf{4.5} and \rf{4.7}, this
case is trivially reduced to the RSII solution. Then, Eqs. \rf{4.7} and \rf{4.9} result in the equation
%%%%%%%
\be{4.27}
2 \La (1+\omega_1)+3(1+\omega_1) b^2+3 \omega_1 b'=0,
\ee
%%%%%%%
with the solution
%%%%%%
\be{4.28} b(\xi)=\sqrt{\cfrac{2}{3}\, \La} \tan \left[\cfrac{1}{3\omega_1}\left(-\sqrt{6\La} (1+\omega_1) \xi+3\sqrt{6\La} C_0\right)\right] \ee
%%%%%%%
where $C_0$ is the constant of integration. Here again the case $\om_1=-1$ is reduced to the \cite{Randall:1999vf} solution with the renormalized bulk
cosmological constant. Therefore, we exclude this case. From this equation we obtain the form of the metric coefficient as
%%%%%%%
\be{4.29}
B(\xi)=- \left\{\cosh\left[m \cfrac{1 + \omega_1}{\omega_1}\, \xi \right]\right\}^{\tfrac{\omega_1}{1 + \omega_1}}\, ,
\ee
%%%%%%
where we, first, took into account the negativeness of the bulk cosmological constant: $(2/3)\La\equiv-m^2<0$, second, restored the $\mathbb Z_2$-symmetry with
respect to the section $\xi=0$ setting $C_0=0$ and, third, normalized $B(\xi)$ in such a way that $B(0)=-1$. Obviously, the normalization condition for $A(\xi)$
should be as follows: $A(\xi)=-B(\xi)$. It is worth noting that, up to trivial numerical prefactor, the metric coefficient has the following asymptotic behaviour:
%%%%%%
\be{4.30} B(\xi) \rightarrow -\exp\left[\mathrm{sign}\left(\frac{\om_1}{1+\om_1}\right)m|\xi|\right], \quad |\xi|\to +\infty\, . \ee
%%%%%%%
Therefore, our spacetime is asymptotically anti-de Sitter with the cosmological constant $\La <0$. The asymptotically decreasing (in absolute value of $\xi$)
solution corresponds to $-1<\om_1<0$.

Let us check other equations from the system \rf{4.5}-\rf{4.9}. Taking into account $a=b$ and \rf{4.9} (or, equivalently, \rf{4.5}), we reduce \rf{4.6} to
%%%%%%%
\be{4.31}
(1 + \omega_0) (2 \La + 3 b^2 + 3 b') =0\, .
\ee
%%%%%%%
If we assume $\omega_0\neq-1$, then its solution is
%%%%%%
\be{4.32}
b(\xi)=\sqrt{\cfrac{2}{3}\, \Lambda_5}\tan \left[\cfrac{\sqrt{6\Lambda_5}}{3}\,(C_2-\xi)\right]\, .
\ee
%%%%%%%
This solution cannot be set equal to \rf{4.28} by adjusting the integration constant. Hence, the only way to make the system of field equations consistent is to
put $\omega_0=-1$. Then, we can check that Eq. \rf{4.8} is automatically satisfied in this case (if $b(\xi)$ satisfies \rf{4.27}).

Finally, from \rf{4.9}  we find that the perfect fluid energy density $\varepsilon(\xi)$ has the form
%%%%%%
\be{4.33}
\varepsilon(\xi)=\cfrac{\La}{\omega_1}\left[{\rm sech}\left(m\cfrac{1+\omega_1}{\omega_1}\, \xi\right)\right]^2\, .
\ee
%%%%%%
Obviously, in order to restore the dimensionality, we must replace $\La$ by $\kappa\La$. Therefore, this energy density is localized near $\xi=0$ and positive for
negative values of $\La$ and $\om_1$. Hence, we have constructed the thick brane model in the case of the bulk perfect fluid. It is quite reasonable that far from
the brane (i.e. $|\xi|\to +\infty$) we restore the anti-de Sitter spacetime. An example of such thick brane is plotted in Fig.~\ref{plot5} for the parameter
$\om_1=-0.75$. Here, the blue/bottom and green/top lines depict the metric coefficient $B(\xi)$ and the perfect fluid energy density $\varepsilon
(\xi)$, respectively.
%%%%%%%%%
\begin{figure}
  \includegraphics[width=3 in, height=1.7 in]{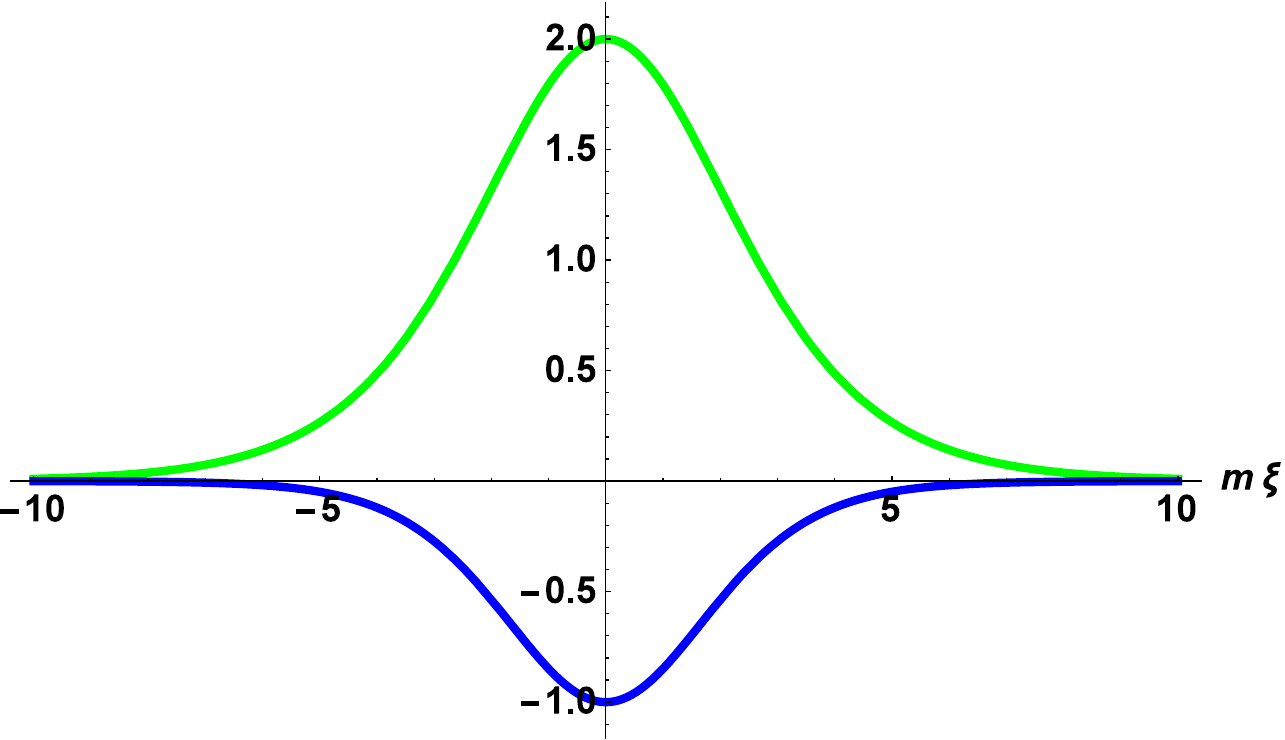}\\
  \caption{The plot of the metric coefficient $B(\xi)=-A(\xi)$ (blue/bottom line) and the dimensionless perfect fluid energy density $(\kappa/m^2)\varepsilon(\xi)$
  (green/top line) in the case of the thick brane.}\label{plot5}
\end{figure}

%%%%%%%%%%%%%%%%%%%%%%%%%%%%%%%%%%%%%%%%%%%%%%%%%%%%%%%%%%%%%%%%%%%%%%%%%%%%%%%%%%%
\section{Conclusion}
\label{sec:conc}

In our paper we considered the static 5D metric with the broken global 4D Poincar\a'{e} invariance. Bulk was filled with the negative cosmological constant and
perfect fluid with anisotropic EoS.

The results of our investigations are twofold. First, we demonstrated that the behavior of models with broken and restored Poincar\a'{e} invariance is
significantly different from each other. Second, our setting of the problem enabled us to obtain new classes of solutions. For example, we have shown that in the case of the empty bulk (the perfect fluid is absent) the solution always has the singularity (naked or coordinate) in contrast to the usual Poincar\a'{e} invariant models (e.g., \cite{Randall:1999ee,Randall:1999vf}). Such type of naked singularities is known for the models with a bulk scalar field and restored Poincar\a'{e} invariance \cite{ArkaniHamed:2000eg,Kachru:2000hf,Cohen:1999ia,Forste:2000ps,Forste:2000ft}. In these papers, the singularities are treated as Big Bang or Big Crunch and
they are taken to effectively cut off space. However, we preferred to construct completely regular solutions. Therefore, we introduced the second brane which cuts
off all singular points where the metric coefficients either are infinite or equal to zero. We found the range of parameters which ensure such regular solutions
defined on the compact space.

Then, we turned our attention to the model with the perfect fluid in bulk and obtained the master equation for the metric coefficients in the case of arbitrary
EoS (except for a couple of special cases). In general case of EoS, this equation does not allow to obtain analytic expressions for the metric coefficients. This
equation is useful for numerical studies of the considered brane world models. We presented two physically interesting particular analytic solutions for the
metric coefficients. The first one generalizes the Randall-Sundrum solution with one brane (RSII) to the case of broken Poincar\a'{e} invariance and bulk perfect
fluid. Here, the perfect fluid has the dust-like EoS parameter $\om_1=0$ in the direction of the fifth coordinate and arbitrary EoS parameter $\om_0$ (except
$-1/3$) in three transverse directions. The second analytic solution describes the thick brane with the restored Poincar\a'{e} invariance. For this model, the
perfect fluid has the vacuum-like EoS $\om_0=-1$ in the transverse directions and $\om_1\neq 0,-1$ in the fifth direction. This solution is of interest since far
from the thick brane it goes asymptotically to the anti-de Sitter one.

To conclude our paper, we would like to mention the following. It is clear that the conclusion whether the obtained solutions can be a realistic model of
our Universe or not depends on the localizability of the zero mode on the brane which recovers the 4D gravity. Clearly, to perform such analysis, we should
investigate the linearized perturbations (including Kaluza-Klein modes) of the considered model. This will be the content of the next paper.

\

%%%%%%%%%%%%%%%%%%%%%%%%%%%%%%%%%%%%%%%%%%%%%%%%%%%%%%%%%%%%%%%%%%%%%%%%%%%

%\vspace{0.5cm}

\section*{ACKNOWLEDGEMENTS}

\"{O}A acknowledges the support by the Distinguished Young Scientist Award BAGEP of the Science Academy. AZ  acknowledges financial support from The Scientific
and Technological Research Council of Turkey (TUBITAK) in the scheme of Fellowships for Visiting Scientists and Scientists on Sabbatical Leave (BIDEB 2221). AZ
also acknowledges the hospitality of \.{I}stanbul Technical University (ITU) where parts of this work were carried out.

%%%%%%%%%%%%%%%%%%%%%%%%%%%%%%%%%%%%%%%%%%%%%%%%%%%%%%%%%%%%%%%%%%%%%%%%%%%%%%%%%%%%%%%%%%%%%%%%%%%%%%%%%%%%%%%%%
%%%%%%%%%%%%%%%%%%%%%%%%%%%%%%%%%%%%%%%%%%%%%%%%%%%%%%%%%%%%%%%%%%%%%%%%%%%%%%%%

\end{document}